# Short-Acquisition Contrast-Free Super-Resolution Microvascular Imaging in Rabbit Kidney

Zhengchang Kou, *Member, IEEE*, Yuning Zhao, *Graduate Student Member, IEEE,* Mingrui Liu, *Graduate Student Member, IEEE,* Rita J. Miller, and Michael L. Oelze, *Senior Member, IEEE*

***Abstract*— Ultrasound localization microscopy (ULM) enables micrometer-scale microvascular imaging by localizing and tracking intravascular microbubbles, but its dependence on exogenous contrast agents and long acquisition times limits clinical translation. This study presents a high-frame-rate contrast-free super-resolution ultrasound microvascular imaging method based on high-frequency ultrafast ultrasound and nonlinear beamforming of backscatter signals from native blood flow. Using only 125 milliseconds of in vivo ultrafast data per image, the proposed method achieved an imaging frame rate of 8 frames/s in a rabbit kidney model. The reconstructed microvascular images resolved vessels with a global spatial resolution of 22.2 μm over a field of view of 23.04 x 15.18 mm², where the wavelength of ultrasound was 67.5 μm. This corresponds to a three-fold improvement over conventional power Doppler imaging under the same acquisition duration. Compared with conventional flow imaging, the proposed method provided improved microvascular contrast and finer vessel delineation without microbubble injection. These results demonstrate a practical pathway toward high frame rate, contrast-free super-resolution ultrasound imaging for microvascular assessment.**

The work of Zhengchang Kou was supported by Beckman Institute Postdoctoral Fellowship. This work was supported in part by the by the National Institutes of Health under grants R01EB036800, R01CA251939, and R01CA273700. (Corresponding author: Michael L. Oelze.)

Zhengchang Kou, Rita J. Miller and Michael L. Oelze are affiliated with the Beckman Institute for Advanced Science and Technology, University of Illinois at Urbana-Champaign, Urbana, IL 61820 USA (email: zkou2@illinois.edu; rjmille@illinois.edu; oelze@illinois.edu); Yuning Zhao, Mingrui Liu, Rita J. Miller and Michael L. Oelze are also with the Department of Electrical and Computer Engineering. In addition, Michael L. Oelze is also with the Carle Illinois College of Medicine and Department of Bioengineering.

## I. Introduction

SUPER-RESOLUTION ultrasound microvascular imaging has rapidly advanced since the emergence of ultrasound localization microscopy (ULM), which breaks the acoustic diffraction limit by localizing and tracking intravascular microbubbles over ultrafast ultrasound imaging sequences [1], [2]. ULM has enabled microvascular imaging at clinically relevant penetration depths and has provided both structural and hemodynamic information that is difficult to obtain with conventional power Doppler or color flow ultrasound. In kidney imaging, ULM is particularly attractive because renal diseases are closely associated with microvascular rarefaction, perfusion impairment, and altered cortical and medullary blood flow. Early studies have shown improved microbubble localization and tracking in rabbit kidney perfusion imaging using spatiotemporal denoising and graph-based microbubble pairing [3], and noninvasive assessment of renal microvascular alterations during acute kidney injury in mice [4]. More recently, high-frame-rate clinical scanners have enabled in-human ULM in organs including the kidney, suggesting a promising route toward clinical translation [5].

Despite these advances, ULM remains fundamentally dependent on the exogenous microbubble contrast agents and typically requires accumulation over an extended acquisition duration to achieve sufficient vascular filling. Long acquisition time, microbubble injection, dependence on robust localization and tracking, computationally expensive post-processing, and sensitivity to physiological or operator-induced motion remain important barriers for routine clinical use, especially in abdominal organs such as the kidney.

To overcome the limitations associated with the contrast-enhanced super-resolution microvascular imaging methods, contrast-free super-resolution microvascular imaging has become an active area of research. Instead of relying on the backscatter signals from microbubbles, the backscatter signals from endogenous red blood cells backscatter are utilized.

Doppler slicing has demonstrated that Fourier-based decomposition enabled sparse representation could enable super-resolution microvascular imaging in mouse brain without contrast agents [6]. Deep-learning-based contrast-free super-resolution Power Doppler (CS-PD) has shown that neural networks can learn a mapping from short-duration contrast-free ultrafast Doppler data to ULM-like microvascular images, achieving improved spatial resolution compared with conventional power Doppler [7]. The neural network approaches demonstrated the potential of learning-based super-resolution from native blood signals, but its performance depends on training data, ULM-derived targets, and cross-domain generalization.

Recently, super-resolution ultrasound imaging using erythrocytes (SURE), which detects and tracks erythrocyte scatterers to generate density and velocity maps without microbubble injection has been proposed [8], [9], [10], [11]. SURE eliminates the need for exogenous contrast agents and can shorten acquisition time from minute to second. However, tracking dense and weak erythrocyte scatterers remains challenging. Besides, motion correction, tissue-signal removal, peak detection, and tracking are still required. More recently, erythrocyte ULM using high-frequency ultrasound was reported for contrast-free cerebrovascular imaging, achieving high spatial resolution from approximately 30 seconds of data acquisition [12]. These studies collectively demonstrate the feasibility of contrast-free super-resolution microvascular imaging, but the acquisition duration remains on the order of seconds or longer in most implementations, and many approaches still rely on localization, tracking, or learned image reconstruction.

As an alternative to localization- or tracking-based methods, nonlinear beamforming can improve the spatial resolution of microvascular imaging directly from ultrafast ultrasound data. We previously developed null subtraction imaging (NSI), a receive-side nonlinear beamforming method that enhances the spatial resolution beyond conventional delay-and-sum beamforming [13], [14]. When combined with ultrafast power Doppler imaging, NSI improved microvascular resolution and contrast while maintaining relatively low computational complexity compared with localization-based super-resolution imaging. This previous work demonstrated that high-resolution power Doppler imaging can be achieved without microbubble localization or tracking, and it established NSI as a promising framework for contrast -free microvascular imaging. However, prior NSI-based power Doppler studies were primarily demonstrated in brain imaging, and the acquisition duration remained on the order of seconds. Whether this nonlinear beamforming framework can be extended to abdominal organs with stronger physiological motion, deeper imaging depth, and shorter data acquisition remains an open question.

In this study, we present an 8-frames/s contrast-free super-resolution microvascular imaging method by extending NSI-based high-resolution power Doppler imaging to *in vivo* rabbit kidney imaging. By integrating high-frequency ultrafast ultrasound acquisition with nonlinear beamforming of endogenous blood-cell backscatter, the proposed method generates one super-resolution microvascular image from only 125 milliseconds of contrast-free ultrafast ultrasound data. This corresponds to an imaging rate of 8 frames/s using non-overlapping acquisition windows. In vivo rabbit kidney experiments demonstrate that the proposed method resolves fine renal microvascular structures over a large field of view without microbubble injection, motion correction, localization, or tracking. Compared with conventional power Doppler imaging, the proposed method provides improved vessel delineation and higher spatial resolution while preserving a short acquisition time. These results suggest that NSI-based contrast-free super-resolution imaging could provide a practical pathway toward clinically translatable microvascular ultrasound imaging of the kidney. In Section II we describe the experiment setup, acquisition parameters, processing pipeline, and image quality evaluation metrics. The *in vivo* results are shown in Section III. Discussion and conclusion of the study are provided in Sections IV and V.

## II. Methods

### A. In Vivo Experiment

The *In vivo* experiment was performed with a 3.5-month-old female New Zealand White rabbit weighing 5.8 kg (Charles River Laboratories, Wilmington, MA). The animal use protocol was approved by the Institutional Animal Care and Use Committee at the University of Illinois at Urbana-Champaign. Anesthesia was induced with 5% isoflurane and maintained with 2% isoflurane, both via face mask. The level of anesthesia was monitored by pedal reflex and respiratory rate. Ophthalmic ointment was applied bilaterally, and the rabbit was placed on a heating pad to maintain body temperature. The skin over the left kidney was shaved. The left kidney was scanned transabdominally by handheld ultrasound probes.

### B. Data Acquisition

TABLE I
Imaging Parameters

| **Probe** | **Visual Sonics MS250** | **Visual Sonics MS200** |
|---|---|---|
| **Transmit Frequency** | 22.73 MHz | 17.86 MHz |
| **Imaging System** | Verasonics NXT 256 High Frequency Configuration | |
| **Pitch** | 0.09 mm | 0.125 mm |
| **Element** | 256 | |
| **Voltage** | 30 V | 35 V |
| **Sampling Frequency** | 90.9091 MHz | 71.4286 MHz |
| **Sample** | 1792 | |
| **Steering Angle** | 9 angles -8 degrees to 8 degrees in 2 degrees step | |
| **PRF** | 2,000 frames per second (18,000 acquisitions per second) | |

Ultrafast ultrasound data were acquired with a Verasonics NXT high-frequency configuration system. To maximize the backscatter signal intensity from the red blood cells (RBCs), we used two high-frequency linear arrays (Visual Sonics MS-200 and Visual Sonics MS-250). Two arrays were driven with transmit frequencies at 17.86 MHz and 22.73 MHz. Table I shows the detailed acquisition parameters. The ultrafast ultrasound data were saved to SSD (Samsung 990 Pro) that was installed on the host computer (Dell Precision 5860).

Both burst and continuous acquisition modes were developed in this study. In the burst acquisition mode, 250 frames were acquired within 0.125 seconds after starting the acquisition, which generated 2 GB of raw channel data. The burst acquisition mode was developed to allow fast acquisition and fast saving. In the continuous acquisition mode, 2,000 frames were acquired within 1 second after starting the acquisition which generated 16 GB of raw channel data. With continuous acquisition mode, 8 non-overlapping acquisition blocks (250 frames) can be partitioned from the 1 second of continuous acquisition.

### C. Processing Pipeline

The ultrafast ultrasound data was first filtered by a singular value decomposition (SVD) filter [15] on the raw channel data [14]. The SVD cutoff was selected empirically. The first 64 and last 64 singular vectors were muted to reject both the tissue signal and noise. Then the NSI beamforming was performed followed by coherent compounding [14]. A fixed DC offset of 0.32 was chosen. After log compression and normalization to 0 dB, the power Doppler images were displayed with automatic dynamic range selection. The dynamic range was set from the mean dB intensity of each image to 0 dB.

The motion correction was not performed in this study because of the ultra-short acquisition time.

The processing pipeline including the SVD filter and NSI beamforming were accelerated with customized CUDA program on GPU (Nvidia RTX 6000 Ada 48GB).

### D. Evaluation Metrics

To provide both qualitative and quantitative evaluation of the image resolution, two evaluation metrics were selected.

First, manually selected cross-sectional profiles were rendered to provide a locally spatial resolution performance comparison between NSI and DAS power Doppler.

Second, the iso-frequency ring was used to provide a global spatial resolution performance evaluation [7]. To estimate the global spatial resolution of NSI power Doppler, we first determined the amplitude cutoff from the iso-frequency ring of DAS power Doppler at the spatial frequency of the acoustic wavelength. The corresponding spatial frequency in the iso-frequency ring of NSI power Doppler at the same amplitude cutoff reflects the spatial resolution of NSI power Doppler.

## III. Results

The *in vivo* results from both ultrasound probes are shown in two separate groups, which are burst acquisition mode and continuous acquisition mode. The images from both DAS power Doppler and NSI power Doppler are displayed side-by-side followed by the cross-sectional profile and iso-frequency ring plot. The display dynamic range is shown in the bottom right corner of each image.

### A. Burst Acquisition Mode

Two planes of the kidney were acquired with a transmit frequency of 22.73 MHz on the MS-250 probe. To minimize the data size and skip the skin and fat layer, the ultrasound data acquisition started at a depth of 3.6 mm. Therefore, the maximum imaging depth with the MS-250 probe in this study reached 18.78 mm.

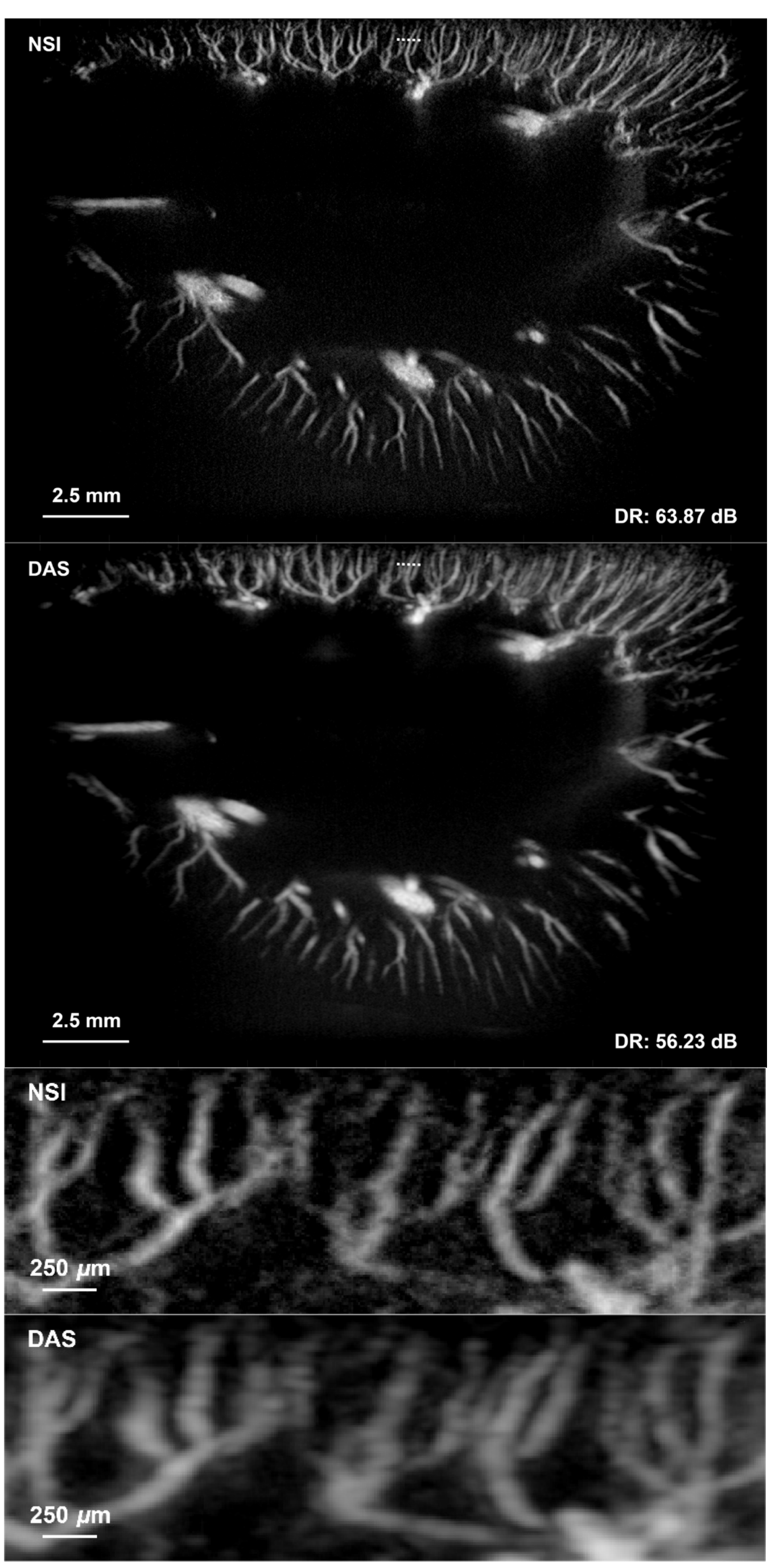

Fig. 1 Contrast-free microvessel images of rabbit kidney with 23 MHz ultrasound array and 0.125 s acquisition duration. Top: NSI and DAS power Doppler images; Bottom: zoomed-in NSI and DAS power Doppler images.

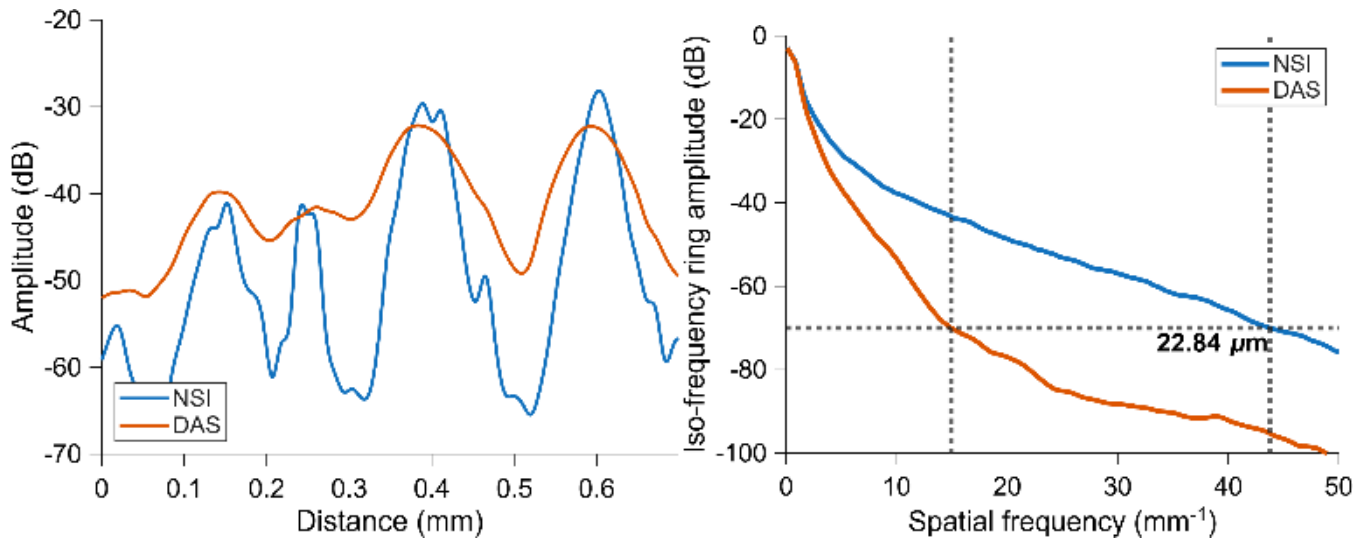

Fig. 2 Local and global resolution comparison between NSI and DAS power Doppler. Left: Cross-sectional profiles of the selected microvessel from both methods. The selected section is marked with dashed line in Fig. 1. Right: Iso-frequency ring curve of both methods. NSI power Doppler achieved a global spatial resolution of 22.84 µm which is 2.97 times better than the acoustic wavelength.

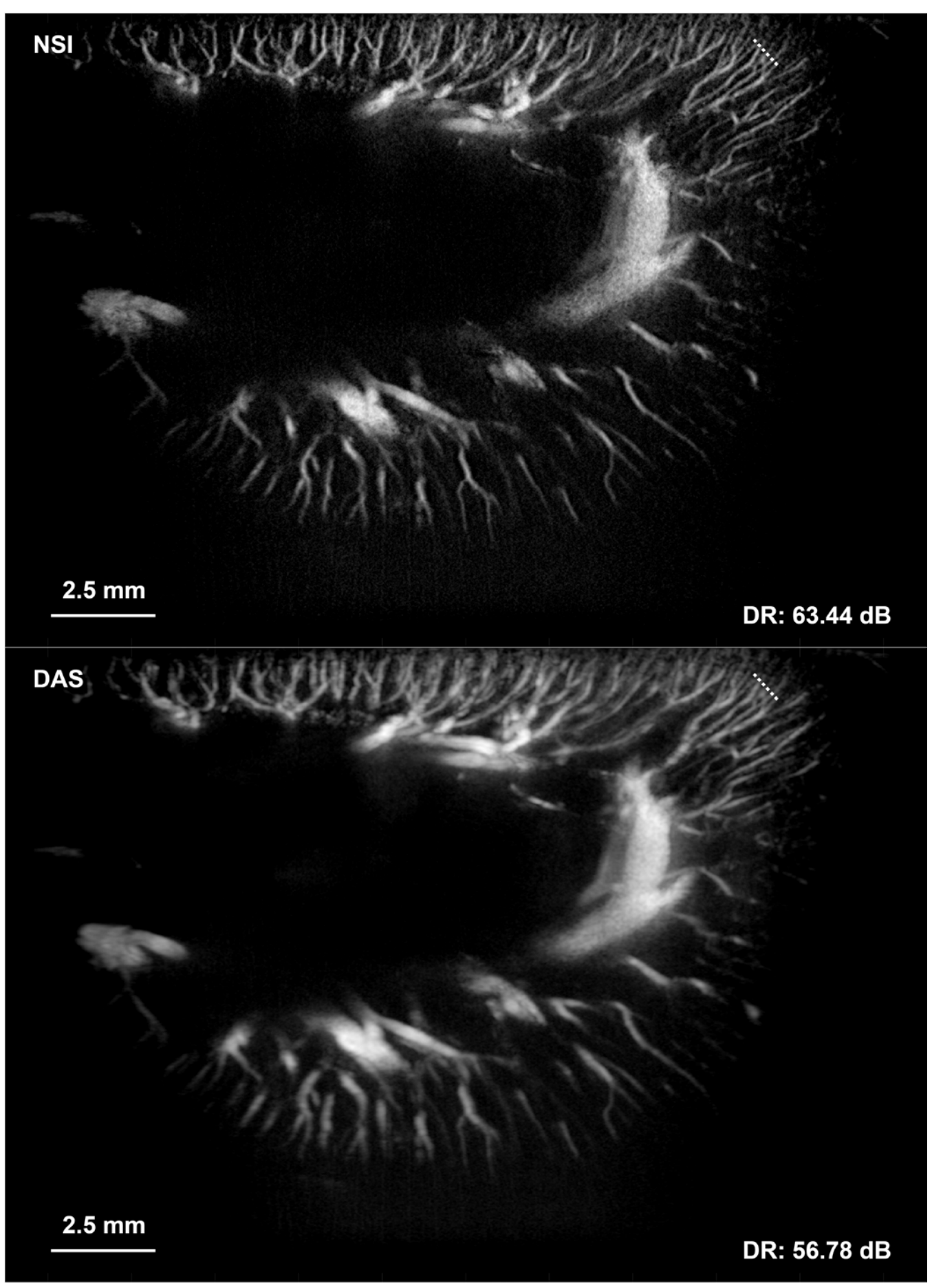

Fig. 3 Contrast-free microvessel images of rabbit kidney with 23 MHz ultrasound array and 0.125 s acquisition duration. Top: NSI power Doppler image. Bottom: DAS power Doppler image.

Figure 1 shows the NSI power Doppler and DAS power Doppler images acquired from the first plane scanned with the MS-250 probe. Spatial resolution improvements can be observed from the NSI power Doppler image even at the lower half of the image. In Fig. 2, the cross-sectional plot of manually selected vessels provides a comparison between the imaging techniques at the top part of the image. NSI successfully resolved vessels that are not resolvable in DAS. From the iso-frequency ring plot in Fig. 2, we estimated a global spatial resolution of 22.84 µm which demonstrates that the NSI power Doppler improved the global spatial resolution by a factor of 2.97 from that of the DAS power Doppler.

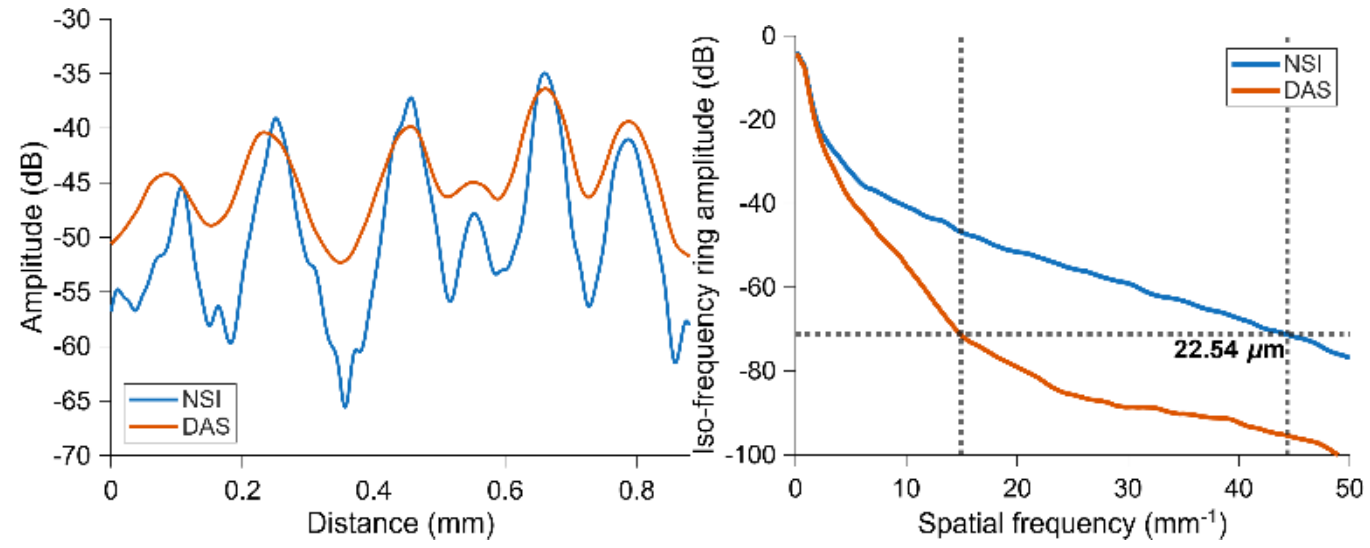

Fig. 4 Local and global resolution comparison between NSI and DAS power Doppler. Left: Cross-sectional profiles of the selected microvessel from both methods. The selected section is marked with dashed line in Fig. 3. Right: Iso-frequency ring curve of both methods. NSI power Doppler achieved a global spatial resolution of 22.54 µm which is 3.01 times better than the acoustic wavelength.

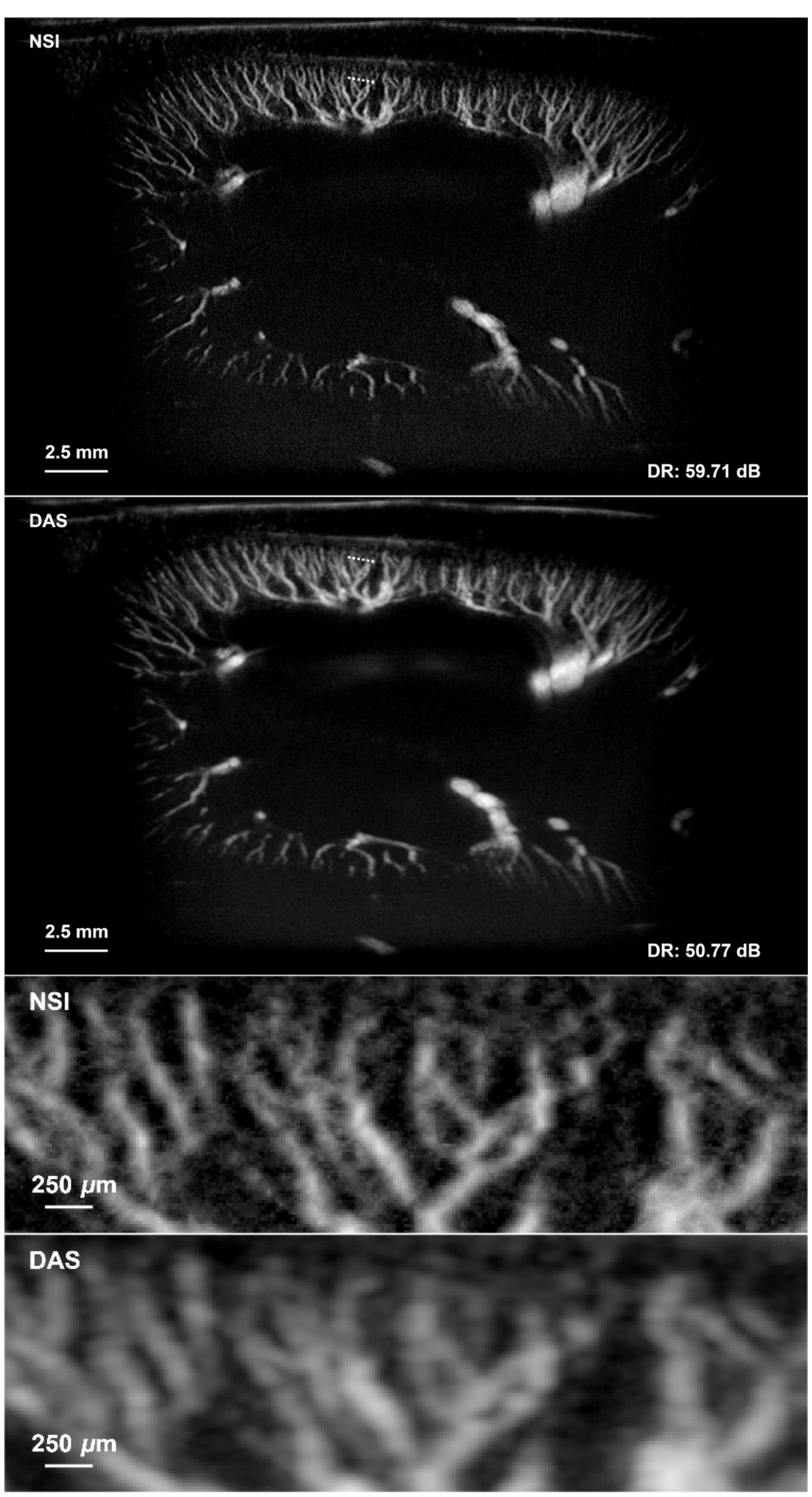

Fig. 5 Contrast-free microvessel images of rabbit kidney with 18 MHz ultrasound array and 0.125 s acquisition duration. Top: NSI and DAS power Doppler images; Bottom: zoomed-in NSI and DAS power Doppler images.

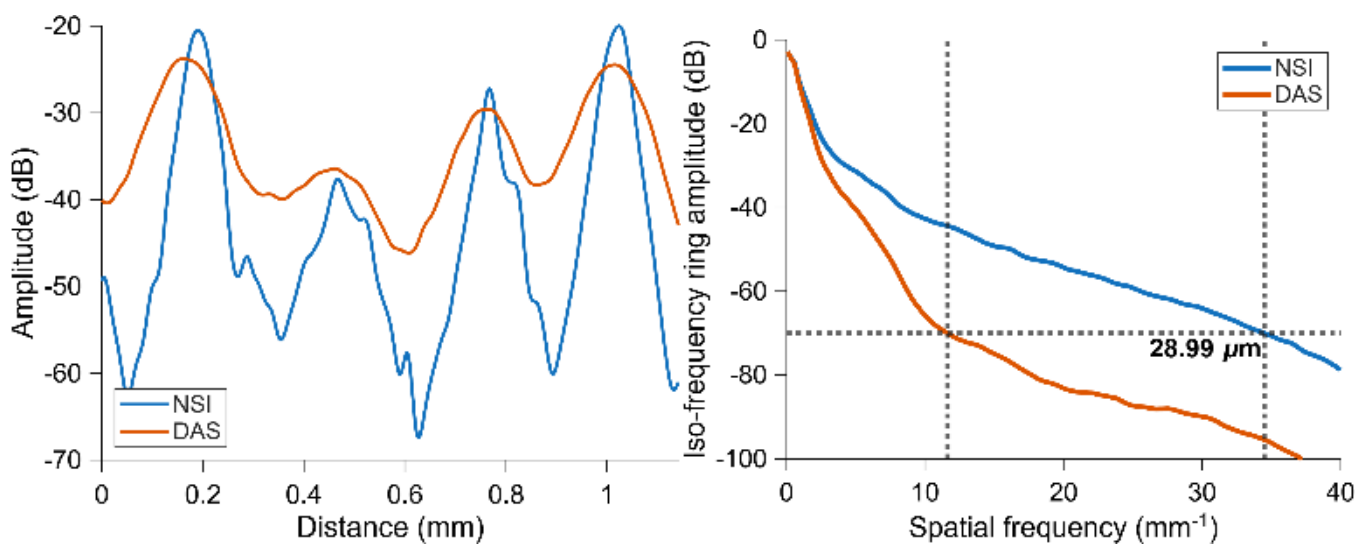

Fig. 6 Local and global resolution comparison between NSI and DAS power Doppler. Left: Cross-sectional profiles of the selected microvessel from both methods. The selected section is marked with dashed line in Fig. 5. Right: Iso-frequency ring curve of both methods. NSI power Doppler achieved a global spatial resolution of 28.99 μm which is 2.97 times better than the acoustic wavelength.

Figure 3 shows the NSI power Doppler and DAS power Doppler images acquired from the second plane scanned with the same probe. The cross-section plot and iso-frequency ring plot in Fig. 4 demonstrate similar performance improvements observed from the first plane that was scanned. From the iso-frequency ring plot in Fig.4, we estimated a global spatial resolution of 22.54 μm, which demonstrates that the NSI power Doppler improved the global spatial resolution by a factor of 3.01 from that of the DAS power Doppler.

Another two planes of the kidney were acquired with a transmit frequency of 17.86 MHz on the MS-200 probe. The ultrasound data acquisition started at a depth of 3.2 mm. Therefore, the maximum imaging depth with the MS-200 probe in this study reached 22.52 mm.

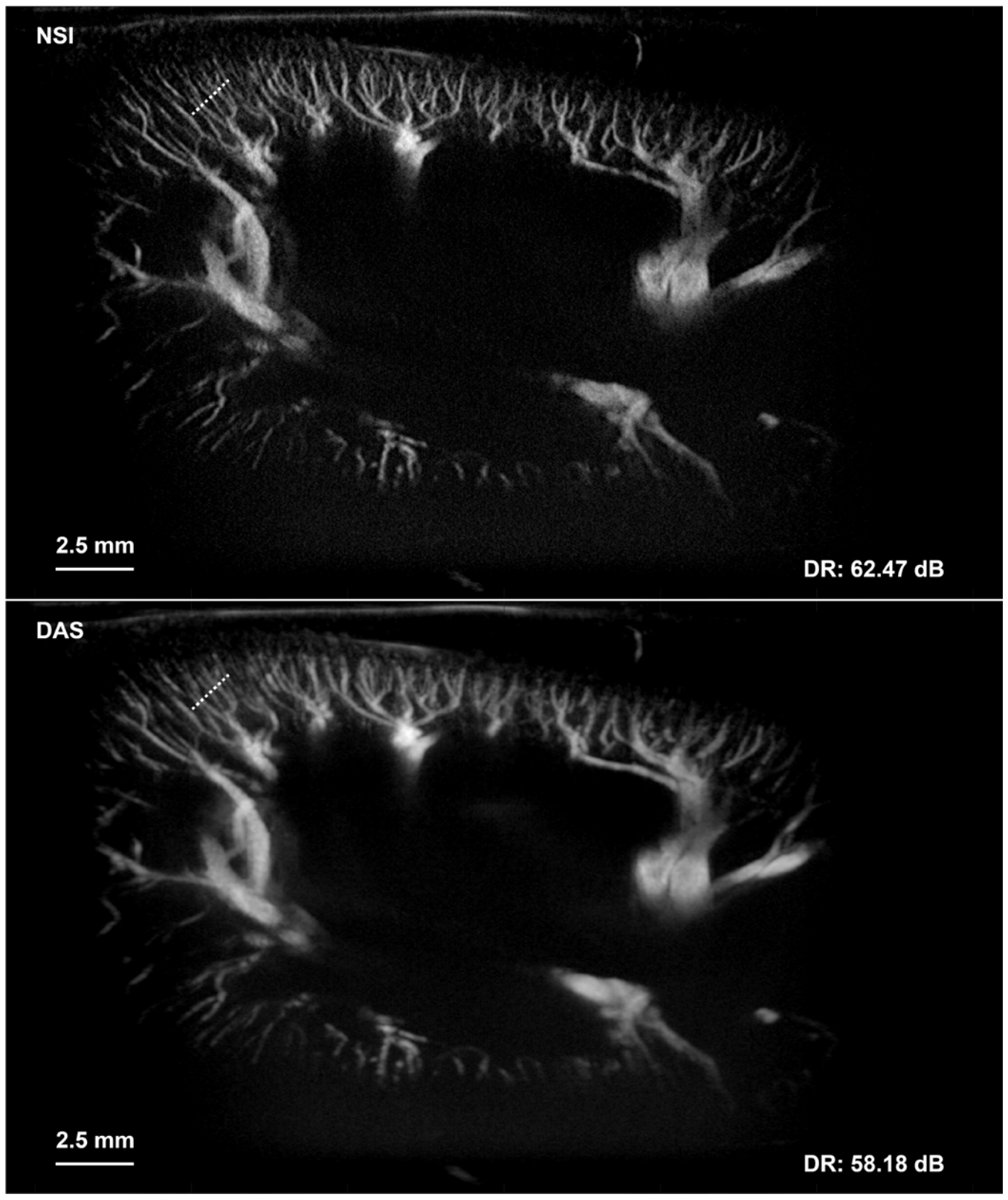

Fig. 7 Contrast-free microvessel images of rabbit kidney with 18 MHz ultrasound array and 0.125 s acquisition duration. Top: NSI power Doppler image. Bottom: DAS power Doppler image.

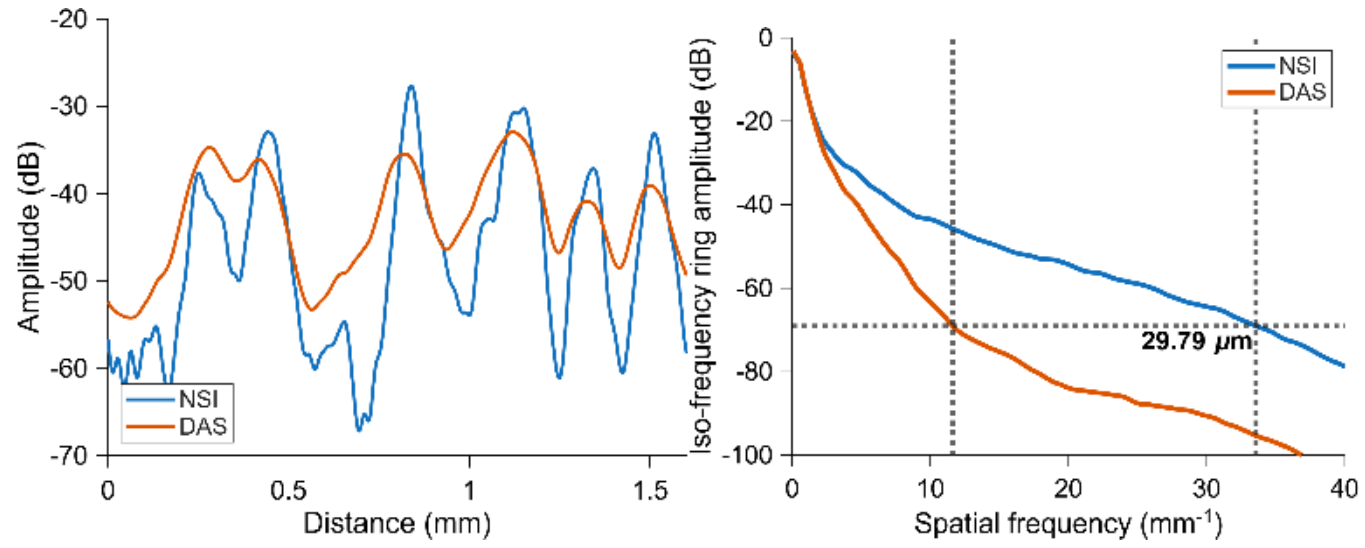

Fig. 8 Local and global resolution comparison between NSI and DAS power Doppler. Left: Cross-sectional profiles of the selected microvessel from both methods. The selected section is marked with dashed line in Fig. 5. Right: Iso-frequency ring curve of both methods. NSI power Doppler achieved a global spatial resolution of 29.79 μm which is 2.89 times better than the acoustic wavelength.

Figure 5 shows the NSI power Doppler and DAS power Doppler images acquired from the first plane scanned with the MS-200 probe. Figure 6 provides the cross-section plot and iso-frequency ring plot. From the iso-frequency ring plot, we estimated a global spatial resolution of 28.99 μm, which demonstrates that the NSI power Doppler improved the global spatial resolution by a factor of 2.97 from that of the DAS power Doppler.

Figure 7 shows the NSI power Doppler and DAS power Doppler images acquired from the second plane scanned with the MS-200 probe. The cross-section plot and iso-frequency ring plot in Fig. 8 demonstrated similar performance improvements observed rom the first plane that was scanned. We estimated a global spatial resolution of 29.79 μm from the iso-frequency ring plot in Fig. 8, which demonstrates that the NSI power Doppler improved the global spatial resolution by a factor of 2.89 from that of the DAS power Doppler.

### *B. Continuous Acquisition Mode*

To move towards video rate contrast-free super-resolution microvascular imaging and further demonstrate the feasibility and stability of the short acquisition contrast-free imaging, we reconstructed eight consecutive and non-overlapping NSI power Doppler images from 2,000 frames that were acquired continuously in 1 second. In the continuous acquisition mode, only the MS-250 probe was used. The acquisition started at a depth of 2.4 mm. Therefore, the maximum imaging depth with the MS-250 probe in the continuous acquisition mode was 17.58 mm. Two planes were acquired with the continuous acquisition mode.

The NSI power Doppler and DAS power Doppler images acquired from the first plane with MS-250 probe are shown in Fig. 9. Each image was reconstructed from 250 frames of ultrafast ultrasound data acquired within 0.125 seconds. The end time of the data acquisition is marked at the bottom left corner of each image.

Figure 10 shows the cross-sectional profiles of manually selected vessels from all the images, and Fig. 11 shows the iso-frequency ring plots of all the images. A mean global spatial resolution of 22.20 μm was estimated from eight sets of iso-frequency ring plots, which demonstrates that the NSI power Doppler improves the global spatial resolution by a factor of 3.05 from that of the DAS power Doppler.

The NSI power Doppler and DAS power Doppler images acquired from the second plane acquired with the continuous acquisition mode are shown in Fig. 12. Figure 13 shows the cross-sectional profiles and Fig. 14 shows the iso-frequency ring plots. A mean global spatial resolution of 22.38 µm was estimated, which demonstrates that the NSI power Doppler improves the global spatial resolution by a factor of 3.03 from that of the DAS power Doppler.

The consistent performance from all the images in two planes demonstrates the stability of the NSI power Doppler with ultra-short acquisition and its robustness against the motion.

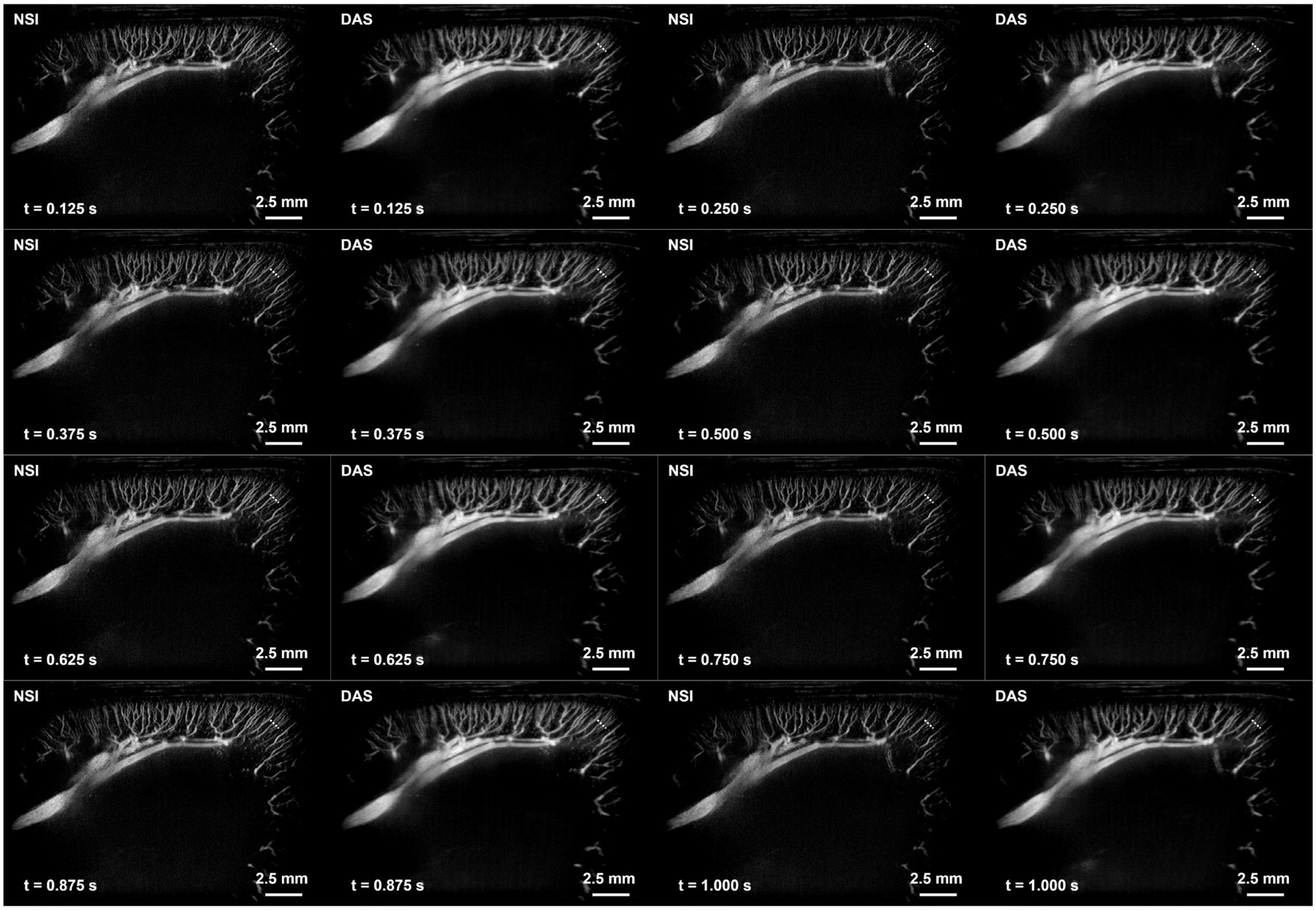

Fig. 9 Contrast-free microvessel images of rabbit kidney with 23 MHz ultrasound array. 2,000 frames were acquired within 1 second. Each image was reconstructed from consecutive blocks of 250 frames (0.125 s acquisition). Both NSI and DAS power Doppler images are shown for side-by-side comparison.

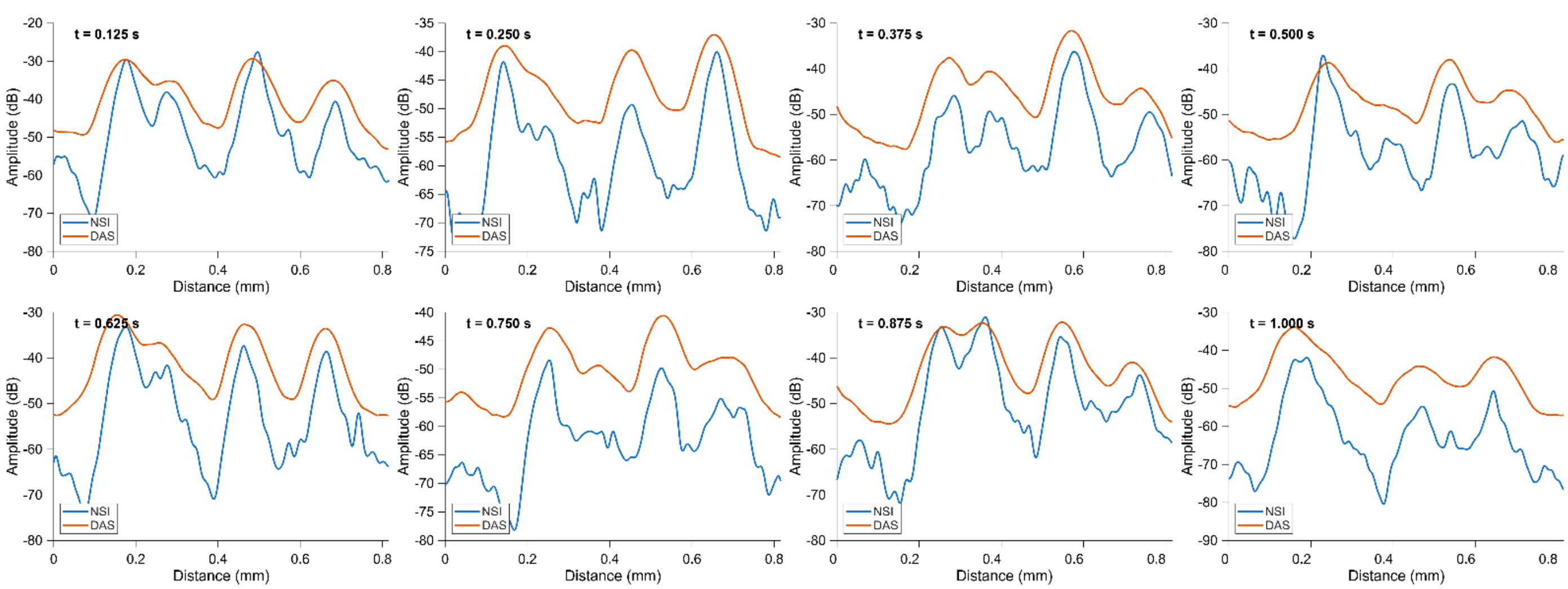

Fig. 10 Local resolution comparison between NSI and DAS power Doppler along 8 consecutive contrast-free microvessel images shown in Fig. 9.

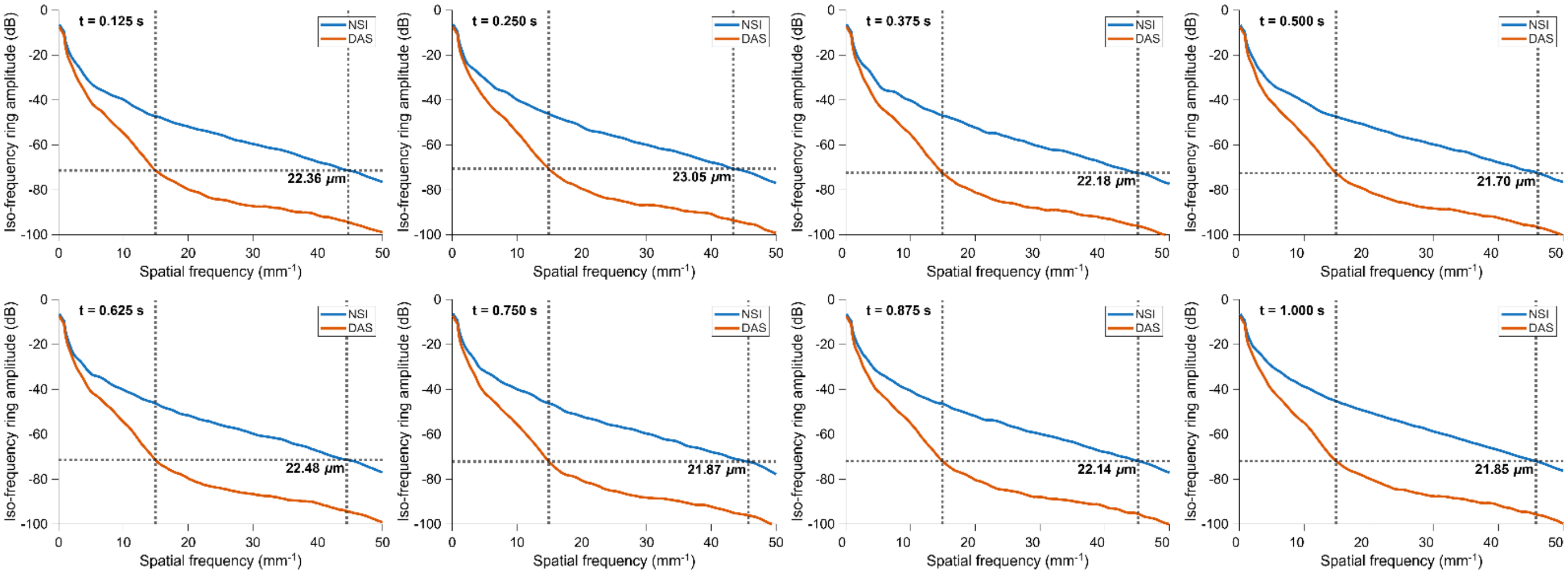

Fig. 11 Global resolution comparison between NSI and DAS power Doppler along 8 consecutive contrast-free microvessel images shown in Fig. 9 using iso-frequency ring. A mean global resolution of 22.20 µm and a standard deviation of 0.4323 µm were measured from these images.

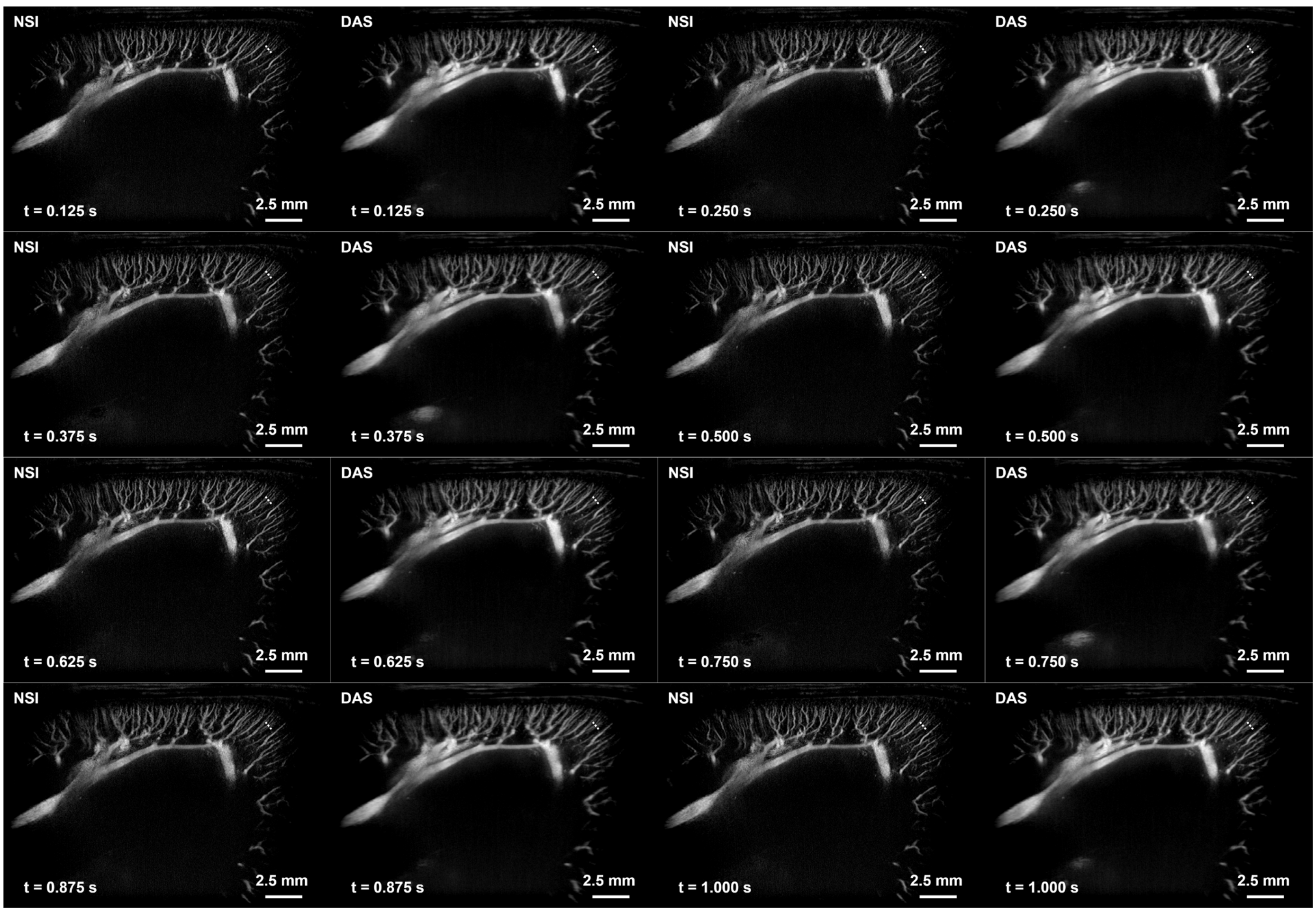

Fig. 12 Contrast-free microvessel images of rabbit kidney with 23 MHz ultrasound array. 2,000 frames were acquired within 1 second. Each image was reconstructed from consecutive blocks of 250 frames (0.125 s acquisition). Both NSI and DAS power Doppler images are shown for side-by-side comparison.

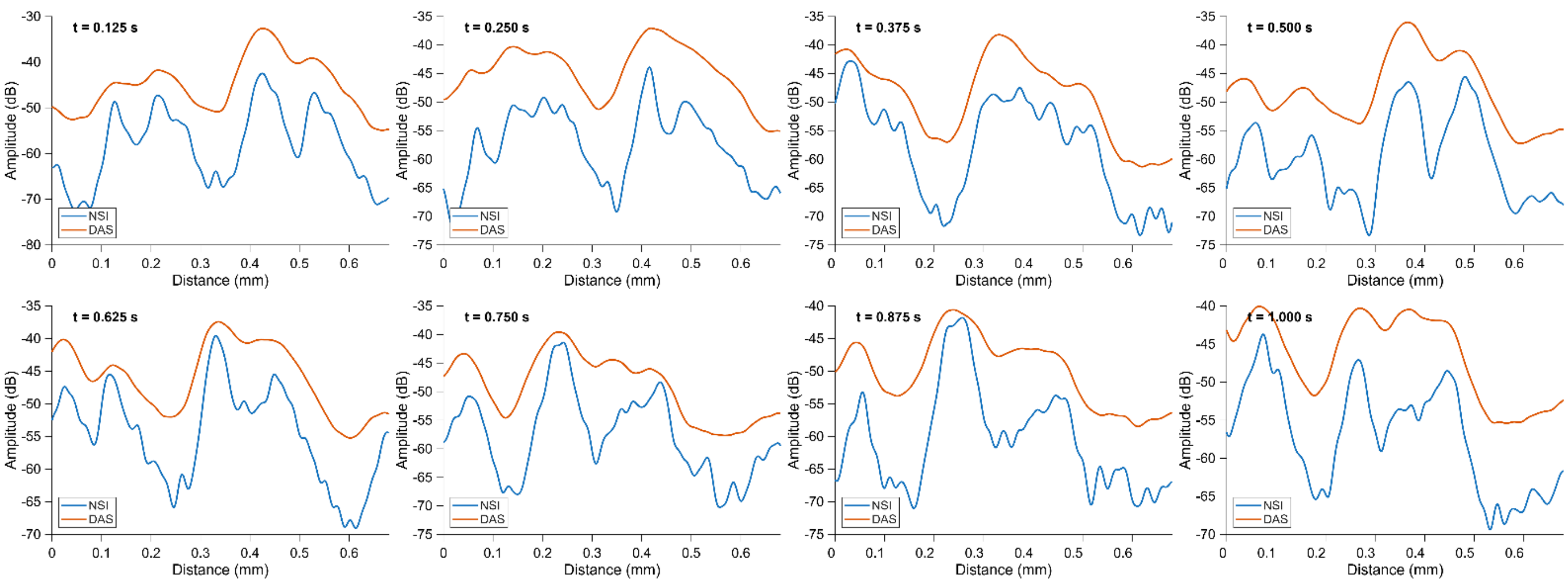

Fig. 13 Local resolution comparison between NSI and DAS power Doppler along 8 consecutive contrast-free microvessel images shown in Fig. 12.

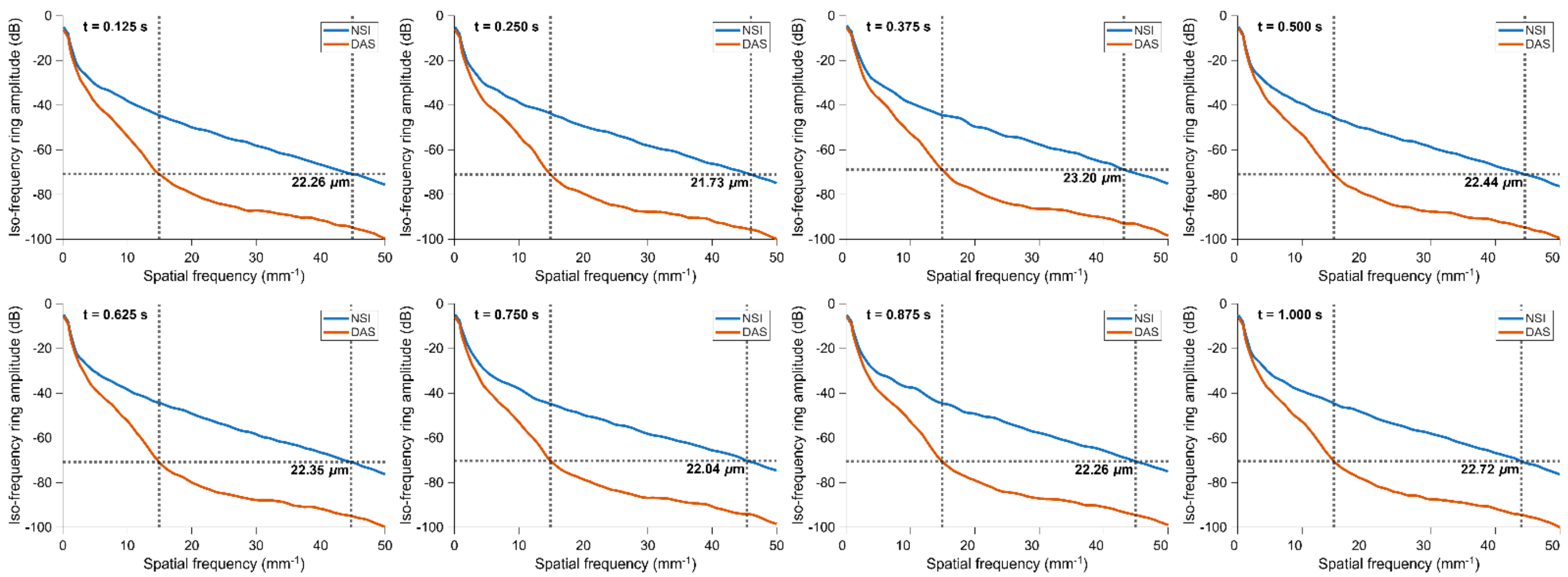

Fig. 14 Global resolution comparison between NSI and DAS power Doppler along 8 consecutive contrast-free microvessel images shown in Fig. 12 using iso-frequency ring. A mean global resolution of 22.38 µm and a standard deviation of 0.4412 µm were measured from these images.

## IV. Discussion

This study demonstrated the feasibility of high-frame-rate contrast-free super-resolution microvascular imaging in rabbit kidney using NSI power Doppler imaging. By combining high-frequency ultrafast ultrasound acquisition with nonlinear receive beamforming, the proposed method generated one super-resolution microvascular image from only 125 milliseconds of native blood-flow data. In the continuous acquisition experiments, eight consecutive and non-overlapping acquisition blocks were reconstructed from 1 s of data, corresponding to an imaging frame rate of eight frames/s. Across two imaging planes acquired with the 23 MHz array, the proposed method achieved mean global spatial resolutions of 22.20 µm and 22.38 µm, corresponding to approximately a three-fold improvement over conventional DAS power Doppler under the same acquisition duration. These results indicate that NSI power Doppler can provide stable contrast-free super-resolution microvascular images in the kidney without microbubble injection, motion correction, localization, or tracking.

The main significance of this work is the substantial reduction in acquisition time for contrast-free super-resolution microvascular imaging. ULM has enabled micrometer-scale vascular imaging by localizing and tracking intravascular microbubbles, but it typically requires exogenous contrast agents and accumulation over extended acquisition durations to achieve sufficient microbubble sampling. These requirements become particularly challenging in abdominal organs, where respiratory motion, probe motion, and physiological motion can degrade localization and tracking accuracy. In contrast, the proposed method uses backscatter signals from RBCs and directly forms high-resolution power Doppler images through nonlinear beamforming. Therefore, it avoids the need for microbubble injections and removes the localization and tracking steps required by ULM. The 125 milliseconds acquisition window also reduces the sensitivity to motion, which is especially important for kidney imaging.

Compared with existing contrast-free super-resolution approaches, the proposed method provides a different balance among resolution, acquisition speed, and processing complexity. Erythrocyte localization methods and SURE have shown that endogenous red blood cells can be used as contrast-free sources for super-resolution ultrasound imaging. However, these methods still rely on scatterer detection and tracking, and the acquisition duration is typically on the order of seconds or longer. Deep-learning-based contrast-free super-resolution power Doppler has also demonstrated promising results by learning a mapping from native power Doppler images to ULM-like vascular maps, but its performance depends on training data, ULM-derived targets, and cross-domain generalization. In comparison, the proposed NSI power Doppler method does not require paired training data. Instead, it enhances spatial resolution at the beamforming stage and can be applied directly to ultrafast ultrasound data.

Kidney imaging is an important and challenging application for contrast-free super-resolution ultrasound. Renal microvascular alterations are associated with many kidney diseases, including acute kidney injury, chronic kidney disease, transplant rejection, and microvascular rarefaction. However, clinical renal ultrasound imaging requires both a large field of view and a short acquisition time. This study demonstrated contrast-free super-resolution microvascular imaging over a large field of view in rabbit kidney. The ability to resolve fine renal microvascular structures from 125 milliseconds acquisition suggests that NSI power Doppler may be suitable for dynamic microvascular assessment, where repeated imaging over time is needed. This capability could be useful for monitoring renal perfusion changes, vascular reactivity, and disease-associated microvascular impairment.

The improvement in spatial resolution arises from the nonlinear beamforming mechanism of NSI. In conventional DAS power Doppler, the spatial resolution is limited by the diffraction-limited receive beam pattern. NSI uses specially designed receive apodizations and nonlinear subtraction to reduce the beamwidth. Because the method operates on native blood-flow signals rather than isolated microbubbles, it is compatible with dense red blood cell backscatter and does not require localization. This feature is particularly important for shortening acquisition time, because the image formation does not depend on accumulating a sufficient number of isolated microbubble or erythrocyte trajectories. The local cross-sectional profiles and global iso-frequency ring analysis consistently demonstrated that NSI power Doppler preserved higher spatial-frequency vascular information than DAS power Doppler.

The continuous acquisition results further support the stability of the method. In two kidney planes, eight consecutive NSI power Doppler images reconstructed from non-overlapping 125 milliseconds acquisition windows showed consistent vessel delineation and global resolution measurements. The standard deviations of the measured global spatial resolution were less than 0.5 μm in both planes, indicating that the resolution improvement was stable across consecutive frames. This is important for potential clinical use because a practical microvascular imaging method should not rely on a single selected acquisition window. Instead, it should maintain consistent image quality across repeated acquisitions.

Several limitations should be acknowledged. First, this study was designed as a feasibility demonstration in an *in vivo* rabbit kidney model. A larger animal cohort and disease models are needed to evaluate reproducibility and biological sensitivity. Second, no independent anatomical ground truth, such as micro-CT, histology, or contrast-enhanced ULM, was used to validate the resolved microvascular structures. Therefore, the reported spatial resolution values should be interpreted as image-based resolution metrics rather than direct estimates of vessel diameter. Third, the iso-frequency ring provided a global estimate of spatial resolution and can be affected by image SNR, vessel orientation, field of view, display dynamic range, and cutoff selection. Future studies should combine iso-frequency analysis with additional quantitative metrics and controlled phantom experiments. Fourth, the SVD clutter-filter cutoff was empirically selected in this study. Adaptive clutter filtering may further improve robustness across different organs, depths, flow speeds, and imaging conditions. Finally, this study used high-frequency arrays to maximize endogenous blood-cell backscatter. Although this is suitable for preclinical kidney imaging and superficial applications, further investigation is needed to evaluate the performance of NSI power Doppler with lower-frequency clinical probes for deeper human abdominal imaging.

Future work will focus on three directions. First, adaptive clutter filtering and automatic parameter selection will be developed to improve robustness and reduce user dependence. Second, validation against contrast-enhanced ULM, histology, or vascular phantoms will be performed to better characterize the accuracy of the resolved microvascular structures. Third, the GPU-accelerated processing pipeline will be further optimized toward real-time implementation. With continued development, high-frame-rate contrast-free NSI power Doppler may provide a practical tool for renal microvascular assessment without contrast-agent injection.

## V. Conclusion

This study demonstrated an eight-frames/s contrast-free super-resolution ultrasound microvascular imaging method for *in vivo* rabbit kidney imaging. By combining high-frequency ultrafast ultrasound with NSI-based nonlinear beamforming, the proposed method generated one super-resolution power Doppler image from only 125 milliseconds of contrast-free ultrafast ultrasound data. The method achieved a global spatial resolution of approximately 22 μm with a three-fold improvement over conventional DAS power Doppler under the same acquisition duration. Stable performance was observed across consecutive non-overlapping acquisition windows, demonstrating the feasibility of high-frame-rate contrast-free renal microvascular imaging. Because the proposed method does not require microbubble injection, motion correction, localization, or tracking, it provides a practical pathway toward clinically translatable super-resolution ultrasound microvascular imaging.